\documentclass[]{emulateapj} % <-- emulador de l'ApJ

\usepackage{epsfig,graphicx,latexsym,amsmath,amssymb}
\usepackage{natbib}
\usepackage{hyperref}
\usepackage{mathrsfs}
\usepackage{lastpage}

\def\msun{M_\odot}
\def\rsun{R_\odot}
\def\mbh{M_\bullet}
\def\t2b{t_{\rm 2b}}
\def\tom{t_\omega}
\def\tRR{t_{\rm RR,s}}
\def\tvRR{t_{\rm RR,v}}

\bibpunct[,]{(}{)}{;}{a}{}{,}

\begin{document}

\author{
Xian Chen\altaffilmark{1}\thanks{e-mail: Xian.Chen@aei.mpg.de} \&
Pau Amaro-Seoane\altaffilmark{1}\thanks{e-mail: Pau.Amaro-Seoane@aei.mpg.de}
}

\altaffiltext{1}{Max Planck Institut f\"ur Gravitationsphysik
(Albert-Einstein-Institut), D-14476 Potsdam, Germany.}

\label{firstpage}

\title{A rapid evolving region in the Galactic Center:\\
       Why S-stars thermalize and more massive stars are missing}

\begin{abstract}
The existence of ``S-stars'' within a distance of $1\arcsec$ from
SgrA$^*$ contradicts our understanding of star formation, due to the
forbiddingly violent environment.  A suggested possibility is that they form
far and have been brought in by some fast dynamical process, since they are
young.  Nonetheless, all conjectured mechanisms either fail to reproduce their
eccentricities --without violating their young age-- or cannot explain the
problem of ``inverse mass segregation'': The fact that lighter stars (the
S-stars) are closer to SgrA$^*$ and more massive ones, Wolf-Rayet (WR) and
O-stars, are farther out.  In this Letter we propose that the responsible for
{\em both}, the distribution of the eccentricities and the paucity of massive
stars, is the Kozai-Lidov-{\em like} resonance induced by a sub-parsec disk
recently discovered in the Galactic center. Considering that the disk probably
extended to smaller radius in the past, we show that in as short as (a few)
$10^6$ years, the stars populating the innermost $1\arcsec$ region would
redistribute in angular-momentum space and recover the observed
``super-thermal'' distribution. Meanwhile, WR and O-stars in the same region
intermittently attain ample eccentricities that will lead to their tidal
disruptions by the central massive black hole. Our results provide new
evidences that SgrA$^*$ was powered several millions years ago by an accretion disk
as well as by tidal stellar disruptions.
\end{abstract}


\maketitle

\section{Introduction}
\label{sec:intro}

Observations of the Galactic Center (GC) going back as far as 20 years
\citep[see][ for a review]{gen10} reveal three facts: (1) An isotropic cusp of
young O/B and Wolf-Rayet (WR) stars, starting at a distance of $30\arcsec$ from
SgrA$^*$ and extending inward to about $1\arcsec$ ($1\arcsec\simeq0.04$ pc);
(2) a mildly thick stellar {\em disk}, of about $100$ WR and O-type stars,
spanning from an inner radius of $1\arcsec$ to an outer radius of about
$10\arcsec$; and (3) populating the innermost region, within $1\arcsec$ from
SgrA$^*$, are a population of B stars, commonly referred to as the ``S-stars'', 
but {\em no} WR/O stars.  A single star-formation (SF)
episode may have explained the formation of disk and cusp stars
\citep[][]{lu13}, the S-stars, however, cannot have been born in this scenario,
because the violent environmental conditions within $1\arcsec$ do not allow
in-situ SF.  One way of populating that region is by dynamical friction, but
the associated timescale is too long. For this reason, the problem has the
reputation of the ``paradox of youth'' \citep{morris93,ghez03}.

This issue has led to the idea that S-stars could have formed at larger radii
and brought in later by an efficient dynamical mechanism. One possibility is
tidal separation of binaries \citep{Hills1991,gould03,gin06}.  A binary, formed
at larger radius, can be set in such an orbit that at periapsis it will be
tidally separated by the central massive black hole (MBH), leaving one star,
which could be a B star, bound to the MBH at a typical radius of $\la
1\arcsec$.  However, the captured stars would have very high eccentricities,
typically about $0.93-0.99$ (see the original work of \citealt{Hills1991} and
\citealt{ama12} for a review).  It would require some $20-50$ Myr for them to
achieve the observed near-thermal distribution
\citep[e.g.][]{perets09,antonini13,zhang13}, with the aid of a very dense cusp
of segregated old stars which is in contradiction with current observations
\citep{buc09,do09}.  Even in the presence of a dense cusp, there are two
additional issues: (1) the same process would work for WR/O stars, and we do
{\em not} see them within $\la 1\arcsec$ \citep{Tal11} and (2) the oldest O/WR
stars are $\la 10$ Myr, so there has to be at least two SF episodes, since
S-stars inside $1\arcsec$ need $\ge 20$ Myr to thermalize.

Since the stellar disk initially must be gaseous, it has been proposed as
another possibility that B stars migrated in it towards the center
\citep{levin07,griv10}. However, (1) this cannot explain the eccentricities of
the S-stars because the migrating stars will remain on near-circular orbits
\citep{perets09,madigan11,antonini13}, and (2) WR/O stars would have migrated
towards the center due to the same mecahnism,  but we do not observe them
there.  Once SF is over (no more gas), the stars in the disk, including the
WR/O stars, would secularly torque each other and drift away from
nearly-circular orbits to rather eccentric ones \citep{madigan09}, and hence at
periapsis they would populate the central $1\arcsec$, but, still, we do not see
WR/O-stars there either.

In this Letter we show that, provided the disk was heavier and more extended in
the past \citep{nay07,wardle08,Bon08,hobbs09,alig11,map12}, it created a rapid
evolving region (RER) inside $1\,\arcsec$, where the angular momenta of
stars rapidly redistribute because of a Kozai-Lidov-like resonance. This RER
can explain both the eccentricities of S-stars  and the absence of WR/O-stars
because the latter are tidally disrupted.

\section{Disk-driven evolution}
\label{sec:model}

\subsection{Timescales}

To understand the effect of the disk, we first analyze the
torque exerted by a wire of mass $\delta m$ and radius $R$ on a background star
of semi-major axis $a$ \citep{iva05,subr05,lockmann08}. \cite{chen11} showed
that the timescale for the wire to change the angular momentum of the star by a
full cycle, i.e. to vary the eccentricity $e$ of the star from its minimum
value to the maximum and back, is

\begin{align}
 T_{K} &=  \left\{
  \begin{array}{lll}
  \frac{2}{3\pi}\frac{\mbh}{\delta m}\left(\frac{a}{R}\right)^{-3}P(a),            &{~\rm Kozai-Lidov,~}   a\le R/2, \\
   \frac{16\sqrt{2}}{3\pi}\frac{\mbh}{\delta m}\left(\frac{a}{R}\right)^{1/2}P(a), &{~\rm Non-determ.,~}   a> R/2,
        \end{array} \right.  \label{eqn:TK}
\end{align}

\noindent
where $\mbh=4\times10^6~\msun$ is the mass of the MBH, and
$P(a)=2\pi\left({a^3}/{G\mbh}\right)^{3/2}\simeq1.4\times10^3~\left({a}/{[0.1~{\rm
pc]}}\right)^{3/2}~{\rm yr}$ is the orbital period of the star. The reason for
$R/2$ is a requirement for having all orbits within the radius of the wire,
including the most eccentric ones, $R_{\rm apo}=a\left(1+e\right) \sim 2\,a$,
with $R_{\rm apo}$ the apocenter distance.  Equation~(\ref{eqn:TK}) is a
generalization of the secular Kozai-Lidov (KL) timescale \citep[see][ and
references therehin]{nao13}: (1) In the regime $a\le R/2$ we recover this
well-known secular phenomenon but (2) when $a\ga R/2$, i.e. when stellar orbits
{\em cross} a sphere with a radius of the wire, it provides good approximation
to the non-deterministic (ND), but not necessarily chaotic, evolution of the
stellar orbit.

Admitting that an extended disk is a superposition of wires, one can derive the
corresponding timescale $T'_K$ for the sum of torques to change the orbital
elements of a star in a full cycle \citep{chang09}:

\begin{equation}
1/T'_K=\int_{R_{\rm in}}^{R_{\rm out}}d(1/T_K),\label{eqn:TKprime}
\end{equation}

\noindent
where $R_{\rm in}$ and $R_{\rm out}$ denote the inner and outer radii of the
disk, $d(1/T_K)\propto \delta m=2\pi\Sigma_d(R)RdR$, and $\Sigma_d(R)$ is the
surface density of the disk.  During $T'_K$, when secular evolution
predominates, a star typically oscillates a full cycle between the maximum and
minimum eccentricities, which are predetermined by three orbital parameters,
namely eccentricity, position angle of periapsis ($\omega$), and inclination
angle relative to the disk ($\theta$). At any intermediate stage of that cycle,
the {``instantaneous''} evolution timescale, defined as $t_K(l)\equiv
l/|\dot{l}|$, can be derived from

\begin{equation}
t_K(l)\simeq lT'_K(a)\label{eqn:tK}
\end{equation}

\noindent
\citep[e.g.][]{chang09,chen11}, where $l\equiv\sqrt{1-e^2}$ is the
dimensionless angular momentum and the dot denotes the time derivative.  The
linear dependence on $l$ reflects the coherence of the disk torque during
$t_K(l)$.

The MBH and cusp stars affect the KL-like evolution by perturbing the
orbital parameters $(e,\,\omega,\,\theta)$. We must distinguish two regimes:
(1) At {\em high} $e$, $\omega$ is significantly perturbed, because of the
induced relativistic (GR) precession rate,

\begin{equation}
\dot{\omega}_{\rm GR}=3(G\mbh)^{3/2}/\left(l^2c^2a^{5/2}\right),
\end{equation}

\noindent
with $c$ the speed of light. It may even exceed the KL precession rate,

\begin{equation}
\dot{\omega}_K\simeq2\pi/\left(T'_K\,l\right)
\end{equation}

\noindent
\citep[e.g.][]{chang09}. When this happens, the disk coherence is broken and
the KL cycle quenched, hence it defines a boundary to the region in phase space
where the evolution is driven by the disk.  In some loose sense, this boundary
is analogous to the Schwarzschild barrier in galactic nuclei
\citep{mer11,brem14}.  (2) At {\em low} $e$, the perturbation on $\omega$
originates from the total stellar mass $M_*(a)$ enclosed by the orbit. The
Newtonian precession rate

\begin{equation}\label{eqn:omegaM}
\dot{\omega}_M\simeq2\pi lM_*(a)/[\mbh P(a)]
\end{equation}

\noindent
may exceed $\dot{\omega}_K$ in this regime, which imposes a second boundary
\citep{chen11}.

Outside these boundaries, evolution of angular-momentum will
be determined by either two-body relaxation, with a characteristic timescale
of

\begin{equation}\label{eqn:t2b}
\t2b(l)\equiv|l/\dot{l}|\simeq l^2\left(\mbh/m_*\right)^2P(a)/\left(N\ln\Lambda\right)
\end{equation}

\noindent
\citep[e.g.][]{koc11}, or (scalar) resonant relaxation \citep[RR, ][]{rau96},
on a timescale of

\begin{align}\label{eqn:tRR}
\tRR(l)\equiv\left|\frac{l}{\dot{l}}\right|\simeq
\frac{l^2}{1-l^2}\left(\frac{\mbh}{m_*}\right)^2\frac{P^2(a)}{N\tom}
\end{align}

\noindent
\citep[][who studied the dependence on $e$]{gur07}.  In
Equations~(\ref{eqn:t2b}) and (\ref{eqn:tRR}), $m_*$ denotes the average mass
of one star,  $N=M_*(a)/m_*$ is the number of stars enclosed by the stellar
orbit, $\ln\Lambda=\ln(\mbh/m_*)$ is the Coulomb logarithm, and
$t_\omega=2\pi/|\dot{\omega}_M-\dot{\omega}_{\rm GR}-\dot{\omega}_K|$ is the
joint precession timescale combining Newtonian, GR, and KL precessions
\citep{chen13}.

Between the two boundaries is the RER: Any star in it cycles between the
maximum and minimum eccentricities predetermined by $(e,\,\omega,\,\theta)$.
Moreover, the two extrema are evolving. The corresponding timescale is given by
vectorial RR \citep{rau96},  which changes $\theta$ on a timescale of

\begin{align}\label{eqn:tvRR}
\tvRR\equiv\left|\frac{1}{\dot{\theta}}\right|
\simeq \frac{0.3}{(0.5+e^2)^{2}}\frac{\mbh}{m_*} \frac{P(a)}{\sqrt{N}}
\end{align}

\noindent
\citep{gur07,eilon09}. Inside RER, $\tvRR$ is longer than the Newtonian and GR
precession timescales, so vectorial RR does not impact the boundaries.  Its
role is to characterize the required time for a star to explore in a
random-walk-fashion the range of maxima and minima in eccentricities fenced in
by the boundaries of the RER.

\subsection{A receding disk}
\label{sec:recedingdisk}

The boundaries of the RER are changing because the properties of the disk have
changed during the past ($1-10$) Myr. We can distinguish two stages in the
evolution of the disk.

\begin{itemize}

\item[(1)] An early phase in which  the disk was mostly gaseous and its
inner edge reached the innermost stable circular orbit (ISCO) at about
$6G\mbh/c^2\simeq10^{-6}~{\rm pc}$ \citep{nay05a,levin07}. This disk contained
at least $10^4~\msun$ of gas, to trigger fragmentation and star formation
\citep{nay05a}, and it could have been as massive as $(3-10)\times10^4~\msun$
according to recent simulations \citep{nay07,Bon08,hobbs09,map12}.  We will
adopt a disk mass of $M_d=3\times10^4~\msun$ for this phase.  It is worth
noting that stars formed in the outer disk may migrate inward
\citep{levin07,griv10}, so the disk inside $R=0.04$ pc could contain both gas
and stars.

\item[(2)] Today, after some $(1-10)$ Myr,  the central $0.04$ pc of the
disk is no longer present, because the gas is consumed by either star formation
\citep{nay05a} or black-hole accretion \citep{alex12}, and the stars have had
time to be scattered out of the disk plane due to vectorial RR
\citep{hop06,koc11}.  For this reason, we say the inner edge of the disk has
{\it receded} from the ISCO to the current location of $R_{\rm
in}\simeq1\arcsec\simeq 0.04~{\rm pc}$ \citep[e.g.][]{paumard06}, while the
outer edge is still the same, at $R_{\rm out}\simeq12\arcsec\simeq 0.5~{\rm
pc}$. The present mass of the disk is $M_d=10^4~\msun$
\citep{paumard06,bartko10}.

\end{itemize}

In both situations, we modeled the disk surface density as a power-law of
$\Sigma_d(R)\propto R^{-1.4}$ \citep{bartko10},  which leads to a mass of
$6\times10^3~\msun$ at $10^{-6}~{\rm pc}<R<0.04~{\rm pc}$ in the early phase.
To derive $M_*(a)$ and $N$ in Equations~(\ref{eqn:omegaM})-(\ref{eqn:tvRR}), we
adopted the broken-power-law model from observations \citep{gen10}, whose
density slop is $\gamma=1.3$ for the inner $0.25$ pc. We assumed an average
stellar mass of $m_*\simeq10~\msun$ \citep[also see][ for discussions]{koc11}.
In this model, we have
$\tvRR\simeq1.5\times10^6~(0.5+e^2)^{-2}(a/1\arcsec)^{0.65}~{\rm yr}$ for stars
in the central arcsec of the Galaxy.

\section{Sculpting the Galactic Center}

\subsection{Rapid Evolving Region}
\label{sec:rer}

\begin{figure*}
\plotone{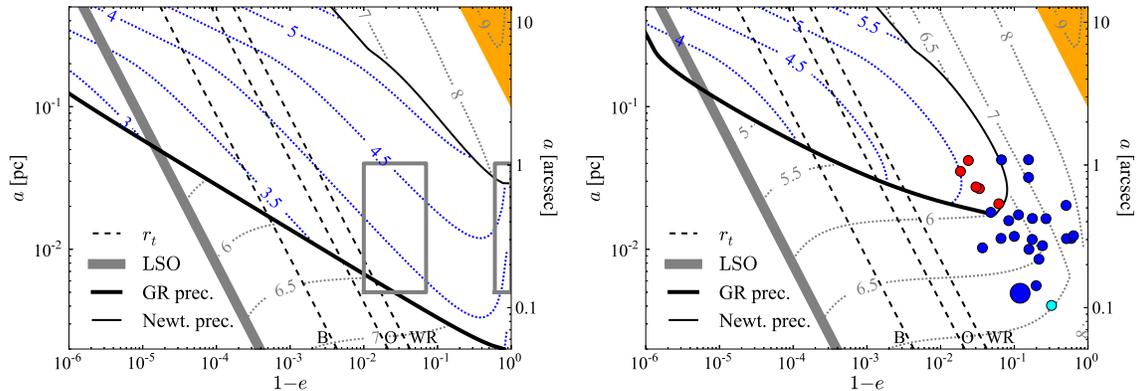}
%\resizebox{\hsize}{!}
%          {\includegraphics[scale=1,clip]{fig1.pdf}}
\caption
   {Mapping evolution timescales in the $a$ -- $(1-e)$ plane.  The thick grey
line on the left-hand-side corresponds to the last stable orbit (LSO) around
SgrA$^*$.  The thick solid black curve is the result of equating the KL
precession rate, $\dot{\omega}_K$, to its relativistic equivalent,
$\dot{\omega}_{\rm GR}$.  Above that curve, dynamical evolution is determined
by KL effect up to the next solid black curve, which comes from equating
$\dot{\omega}_K$ to $\dot{\omega}_M$, the precession rate induced by the
enclosed stellar mass.  The dashed black parallel lines crossing the figures
from the top to the bottom indicate the typical tidal-disruption radii for B,
O, and WR stars.  The blue dotted isochrones, fenced in the region where the
evolution is governed by the KL mechanism, are associated with the logarithms
of the KL timescales given by Equation~(\ref{eqn:thori}).  The grey dotted
isochrones are associated with the logarithms of the two-body-relaxation or RR
timescales, whichever is shorter.  The small orange triangles at the top-right
corners depict the loci of the red giants in the GC. In the left panel, the two
grey boxes depict the expected birth places of S-stars in the binary-separation
and migration-in-disk models \citep[also see][]{antonini13}.  In the right
panel, the dots correspond to: S-stars not associated with the young stellar
disk \citep[small-blue][]{gil09}, the infalling G2 object \citep[also called DSO, see
][ for a different interpretation of its nature]{eckart13} measured at different
times or different wavelengths \citep[small-red][]{gil13rev}, S2/S0-2
\citep[big-blue, the brightest S-star,][]{ghez03,eis03}, and S102/S0-102
\citep[small-cyan,][]{mey12}, the S-star with the shortest period known.}
\label{fig.1} \end{figure*}

In Figure~\ref{fig.1} we display the boundaries of the RER. The left panel
corresponds to $R_{\rm in}=10^{-6}~{\rm pc}$ and the right one to $R_{\rm
in}=0.04~{\rm pc}$.  In this $(1-e)$ -- $a$ plane, at any location, 
we can estimate the instantaneous evolution timescale as:

\begin{equation}\label{eqn:thori}
\left|\frac{1-e}{\dot{e}}\right|=\frac{e(1-e)}{l^2}t_K(l)\simeq \frac{e(1-e)}{l}T'_K(a).
\end{equation}

We can then identify the lines with constant evolution timescales, i.e. the
contours. We call them ``isochrones'', and depict them as blue dotted curves.
Outside the RER the isochrones are shown in grey and determined by either
two-body scattering, $e\,\left(1-e\right)\,t_{\rm 2b}\,(l)/l^2$,  or RR
process, $e\,\left(1-e\right)\,\tRR(l)/l^2$, whichever timescale is shorter.

Two striking conclusions from a first look at this figure are  (1) stars in the
RER evolve on very short timescales, of the order of $10^{3-5.5}$ yrs, to
complete a full KL cycle.  As discussed previously, after a time of $\tvRR$,
any star at $a<1\arcsec$ would have {\em fully} explored the angular-momentum
range within the RER. (2) As the disk recedes, the boundaries come closer and
the RER shrinks.  Any star that finds itself out of the RER will be ``frozen''
from the point of view of another star which still is in it: The timescales
outside RER are long.

Since short evolution timescale leads to low probability of stellar
distribution, today (right panel of Figure~\ref{fig.1}), the absence of S-stars
within the RER boundaries may be a plausible observational corroboration that
the RER does exit at the GC. At present, the only measured object within the
RER boundaries is G2 \citep[red dots, ][]{gil13rev}.  From its nearby
isochrones, we see that G2 must have been formed less than $10^{5.5}$ years
ago.

\subsection{A close thermalization of the S-stars}\label{sec:sstar}

The two more successful scenarios of depositing B stars close to the GC, i.e.
binary separation and disk migration, place these stars well within the RER
(left panel of Figure~\ref{fig.1}).  These stars are able to sufficiently
mix in angular-momentum space, on a timescale of $\tvRR\simeq0.7$ Myr in the
binary-separation scenario and of $\tvRR\simeq6$ Myr in the disk-migration one.
The latter mechanism (disk migration) requires much longer time to reach the
superthermal distribution in eccentricities becasue of the $e$ dependence of
Equation~(\ref{eqn:tvRR}). We note that these timescales are at least $10$
times shorter than those from the earlier models, which neglected the RER.

The fully-mixed eccentricities do not necessarily have a thermal distribution
(Brem et al., in preparation).  Following the argument that longer evolution
timescale correlates with higher probability distribution, we will have $dN/de
\propto dt/de$, and substituting Equation~(\ref{eqn:thori}) for $dt/de$, we can
derive $dN/de \propto dt/de\propto e/l$.  This distribution function is steeper
than a thermal one, $dN/de\propto e$. The steepness stems from the linear
dependence of the evolution timescale ($l/\dot{l}$) on the orbital angular
momentum $l$,  whereas in the case of two-body relaxation and RR, the evolution
time scales with $1-e^2$.

\begin{figure}
%\resizebox{\hsize}{!}
%          {\includegraphics[scale=1,clip]{fig2.eps}}
\plotone{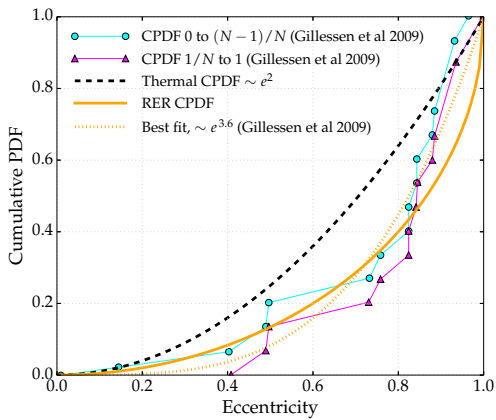}
\caption
   {Cumulative probability distribution functions of the eccentricities of
S-stars, derived from observations (cyan and purple), thermal distribution
($e^2$, black-dashed), our RER model ($1-\sqrt{1-e^2}$, orange-solid), and best
fit to observational data (orange-dotted).} \label{fig.2}
\end{figure}

Figure~\ref{fig.2} compares various cumulative probability distribution
functions (CPDFs) for $e$, derived from two theoretical models --a thermal one
and our RER model-- as well as from observations.  It is clear that compared to
the thermal distribution, the RER one is in better agreement with the
observations. 

\subsection{Depleting WR/O stars}\label{sec:wrostar}

For a star of mass $m_*$ and radius $r_*$, if its eccentricity becomes so high
that the orbital pericenter distance from SgrA$^*$ becomes smaller than the
tidal radius

\begin{equation}
r_t\simeq r_*\left(\frac{\mbh}{m_*}\right)^{1/3}
\simeq4\times10^{-6}~{\rm pc}\left(\frac{r_*}{\rsun}\right)
\left(\frac{m_*}{\msun}\right)^{-1/3},
\end{equation}

\noindent
it will be tidally disrupted \citep{ree88}.  In Figure~\ref{fig.1} we
show typical $r_t$ associated with B, O, and WR stars. Once stars cross it from
the right to the left, they are lost. This drain is enhanced in the RER by the
shorter and shorter timescales, as the stars progressively move to the left.

To calculate $r_t$, we assume $m_*=(7,\,25,\,60)~\msun$ respectively for the
three types of stars \citep{zin07}.  Main-sequence stars less massive than
$7\msun$ are below the current detection limit of observations.
Correspondingly, we have adopted $r_*=4~\rsun$ for main-sequence B stars, and
$r_*=40~\rsun$ and $80~\rsun$ respectively for O- and WR stars. B stars can be
envisaged as main-sequence stars, but O- and WR stars are more massive, and
shorter-lived. We hence adopt larger radii for them, 3--4 times larger than
typical radii on the main-sequence, since they have evolved off the main
sequence \citep{paumard06,bartko10}.

We can see in the left panel of Figure~\ref{fig.1} that any WR star in a stripe
defined between $0.15\arcsec\la a\la0.8\arcsec$ and any O star in
$0.2\arcsec\la a\la0.8\arcsec$ will be tidally disrupted, because it will have
explored all the $(1-e)$ space in $\sim 10^6$ yrs.  Similarly, for B stars, the
corresponding stripe is delimited by the narrower zone $0.5\arcsec\la
a\la0.8\arcsec$. In fact, this predicted gap (for B stars) does occur in the
current distribution of S-stars (right panel of Figure~\ref{fig.1}).  If we
had assumed a disk mass of $M_d>3\times10^4~\msun$, this gap would have
broadened to incorporat the region where $a<0.5\arcsec$, and it would
contradict current observations.  Therefore, an upper limit to the disk mass
can be derived, approximately $3\times10^4~\msun$.

By looking at the left panel again, we realize that only WR/O stars with
$a>0.8\arcsec$ and low $e$ can survive, because they are always outside of the
RER and cannot drift quickly enough to higher $e$.  Indeed, WR/O stars have
been discovered only at $a\ga1\arcsec$ but not inside.  In principle, our
model cannot deplete WR/O stars at $a<0.1\arcsec$, because at such small $a$
the RER does not reach the tidal radii. Observations did not find any WR/O star
there, maybe because the extrapolation of the disk density profile
$\Sigma(R)\propto R^{-1.4}$ results in  $<1$ WR/O star at $R<0.1\arcsec$.

\section{Discussions}\label{sec:dis}

In this Letter we have presented a picture that explains the distribution of
the eccentricities of S-stars and the absence of more massive stars within
$1\arcsec$ of SgrA$^*$.  Our {\em sole} hypothesis is that around $(1-10)$ Myr
ago, the disk had extended down to $R\ll0.04$ pc.  We find that the torque
exerted by the disk creates a region at the GC in which the dynamical evolution
is significantly accelerated as compared to other regions, by a factor ranging
from 10 to 100 times, and we call it the ``rapid evolving region''. 

Our scenario agrees with current observations about the nonexistence of an old
segregated cusp in the GC \citep{buc09,do09}, contrary to other works, which
crucially rely on the cusp to thermalize the S-stars
\citep{perets09,madigan11,antonini13,zhang13}.  Because the time that is
needed to randomize angular momentum is now shortened to $(0.7-6)$ Myr, our
model is able to accommodate various possibilities for the formation of
S-stars, while other models rely heavily on when S-stars were brought to the
GC \citep{perets09,antonini13}.

Moreover, our RER scenario unifies two observational facts that have been
thought until now to be disconnected: We successfully populate the observed
range of $e$ for B stars and we can duplicate the observed discontinuity of
WR/O stars above and below $1\arcsec$. Both of them will be established in as
short as $(0.7-6)$ Myr, so we can even unify the origin of all the young
stellar populations in the GC to {\em only one single SF episode}.  This
unification does pose a problem for earlier models: If all B stars formed
simultaneously with WR/O stars, since this must be less than 6 Myr ago
(because WR/O stars cannot be older),  two-body relaxation and RR will fail
to explain the distribution of $e$.

At this stage, it is crucial to theoretically understand the dynamical
response of the old stellar population to the RER, and test it against the
observations of dimmer (than B-type), older stars.  If they match, it would be
a robust evidence that the RER has indeed played a role in sculpting the GC.

\acknowledgments

This work has been supported by the Transregio 7 ``Gravitational Wave
Astronomy'' financed by the Deutsche Forschungsgemeinschaft DFG (German
Research Foundation). We thank Bence Kocsis, Meng Su, and Scott Tremaine
for discussions, and the Kavli Institute for Theoretical
Physics where one part of this work has been completed. This research was
supported in part by the National Science Foundation under Grant No.  NSF
PHY11-25915.  PAS is indebted with Sonia P{\'e}rez for conversations and
extraordinary support.

\label{lastpage}

\begin{thebibliography}{}

\bibitem[Alexander(2011)]{Tal11} Alexander, T.\ 2011, The Galactic Center: a Window to the Nuclear Environment of Disk Galaxies, 439, 129

\bibitem[Alexander et al.(2012)]{alex12} Alexander, R.~D.,
Smedley, S.~L., Nayakshin, S., \& King, A.~R.\ 2012, \mnras, 419, 1970


\bibitem[Alig et al.(2011)]{alig11} Alig, C., Burkert, A.,
Johansson, P.~H., \& Schartmann, M.\ 2011, \mnras, 412, 469

\bibitem[Amaro-Seoane (2012)]{ama12} Amaro-Seoane, P.\ 2012, Submitted to Living Reviews in Relativity

\bibitem[Antonini \& Merritt(2013)]{antonini13} Antonini, F., \& Merritt, D.\ 2013, \apjl, 763, L10

\bibitem[Bartko et al.(2010)]{bartko10} Bartko, H. \ 2010, \apj, 708, 834

\bibitem[Bonnell \& Rice(2008)]{Bon08} Bonnell, I.~A., \& Rice, W.~K.~M.\ 2008, Science, 321, 1060

\bibitem[Brem et al.(2014)]{brem14} Brem, P., Amaro-Seoane,
P., \& Sopuerta, C.~F.\ 2014, \mnras, 437, 1259

\bibitem[Buchholz et al.(2009)]{buc09} Buchholz, R.~M., Sch{\"o}del, R., \& Eckart, A.\ 2009, \aap, 499, 483

\bibitem[Chang(2009)]{chang09}Chang, P. 2009, \mnras, 393, 224

\bibitem[Chen \& Liu(2013)]{chen13} Chen, X., \& Liu, F.~K.\ 2013, \apj, 762, 95

\bibitem[Chen et al.(2011)]{chen11}Chen, X., Sesana, A., Madau, P., \& Liu, F. K.\ 2011, \apj, 729, 13

\bibitem[Do et al.(2009)]{do09} Do, T., Ghez, A.~M., Morris, M.~R., et al.\ 2009, \apj, 703, 1323

\bibitem[Eilon et al.(2009)]{eilon09} Eilon, E., Kupi, G., \& Alexander, T.\ 2009, \apj, 698, 641

\bibitem[Eisenhauer et al.(2003)]{eis03} Eisenhauer, F.,
Sch{\"o}del, R., Genzel, R., et al.\ 2003, \apjl, 597, L121

\bibitem[Eckart et al.(2013)]{eckart13} Eckart, A., Mu{\v z}i{\'c}, K., Yazici, S., et al.\ 2013, \aap, 551, A18 



\bibitem[Genzel et al.(2010)]{gen10} Genzel, R., Eisenhauer, F., \& Gillessen, S.\ 2010, RvMP, 82, 3121

\bibitem[Ghez et al.(2003)]{ghez03} Ghez, A.~M., Duch{\^e}ne,
G., Matthews, K., et al.\ 2003, \apjl, 586, L127

\bibitem[Gillessen et al.(2009)]{gil09} Gillessen, S., Eisenhauer, F., Trippe, S., et al.\ 2009, \apj, 692, 1075


\bibitem[Gillessen et al.(2013)]{gil13rev} Gillessen, S.,
Genzel, R., Fritz, T.~K., et al.\ 2013, arXiv:1312.4386

\bibitem[Ginsburg \& Loeb(2006)]{gin06} Ginsburg, I., \& Loeb, A.\ 2006, \mnras, 368, 221

\bibitem[Gould \& Quillen(2003)]{gould03} Gould, A., \& Quillen, A.~C.\ 2003, \apj, 592, 935

\bibitem[Griv(2010)]{griv10} Griv, E.\ 2010, \apj, 709, 597

\bibitem[G{\"u}rkan \& Hopman(2007)]{gur07} G{\"u}rkan, M.~A., \& Hopman, C.\ 2007, \mnras, 379, 1083

\bibitem[Hills(1991)]{Hills1991} Hills, J.~G.\ 1991, \aj, 102,704

\bibitem[Hobbs \& Nayakshin(2009)]{hobbs09} Hobbs, A., \& Nayakshin, S.\ 2009, \mnras, 394, 191

\bibitem[Hopman \& Alexander(2006)]{hop06} Hopman, C., \& Alexander, T.\ 2006, \apjl, 645, 1152

\bibitem[Ivanov et al.(2005)]{iva05}Ivanov, P. B., Polnarev, A. G., \& Saha, P., 2005, \mnras, 358, 1361

\bibitem[Kocsis \& Tremaine(2011)]{koc11} Kocsis, B., \& Tremaine, S. 2011, \mnras, 412, 187

\bibitem[Levin(2007)]{levin07} Levin, Y.\ 2007, \mnras, 374, 515

\bibitem[Levin \& Beloborodov(2003)]{levin03} Levin, Y., \& Beloborodov, A.~M.\ 2003, \apjl, 590, L33 


\bibitem[L{\"o}ckmann et al.(2008)]{lockmann08} L{\"o}ckmann, U.,
Baumgardt, H., \& Kroupa, P.\ 2008, \apjl, 683, L151

\bibitem[Lu et al.(2013)]{lu13} Lu, J.~R., Do, T., Ghez, A.~M., et al.\ 2013, \apj, 764, 155

\bibitem[Madigan et al.(2011)]{madigan11} Madigan, A.-M., Hopman,
C., \& Levin, Y.\ 2011, \apj, 738, 99

\bibitem[Madigan et al.(2009)]{madigan09} Madigan, A.-M., Levin,
Y., \& Hopman, C.\ 2009, \apjl, 697, L44

\bibitem[Mapelli et al.(2012)]{map12} Mapelli, M., Hayfield, T., Mayer, L., \& Wadsley, J.\ 2012, \apj, 749, 168

\bibitem[Merritt et al.(2011)]{mer11} Merritt, D., Alexander, T., Mikkola, S., \& Will, C.~M.\ 2011, \prd, 84, 044024

\bibitem[Meyer et al.(2012)]{mey12} Meyer, L., Ghez, A.~M., Sch{\"o}del, R., et al.\ 2012, Science, 338, 84

\bibitem[Morris(1993)]{morris93} Morris, M.\ 1993, \apj, 408, 496

\bibitem[Naoz et al.(2013)]{nao13} Naoz, S., Farr, W.~M., Lithwick, Y., Rasio, F.~A., \& Teyssandier, J.\ 2013, \mnras, 431, 2155

\bibitem[Nayakshin \& Cuadra(2005)]{nay05a} Nayakshin, S., \& Cuadra, J.\ 2005, \aap, 437, 437

\bibitem[Nayakshin et al.(2007)]{nay07} Nayakshin, S.,  Cuadra, J., \& Springel, V.\ 2007, \mnras, 379, 21

\bibitem[Paumard et al.(2006)]{paumard06}Paumard, T., et al.\ 2006, \apj, 643, 1011

\bibitem[Perets et al.(2009)]{perets09} Perets, H.~B., Gualandris, A., Kupi,
G., Merritt, D., \& Alexander, T.\ 2009, \apj, 702, 884

\bibitem[Rauch \& Tremaine(1996)]{rau96}Rauch, K. P., \& Tremaine, S. 1996, NewA, 1, 149

\bibitem[Rees(1988)]{ree88}Rees, M. J.\ 1988, \nat, 333, 523

\bibitem[{\v S}ubr \& Karas(2005)]{subr05} {\v S}ubr, L., \& Karas, V.\ 2005, \aap, 433, 405

\bibitem[Wardle \& Yusef-Zadeh(2008)]{wardle08} Wardle, M., \& Yusef-Zadeh, F.\ 2008, \apjl, 683, L37

\bibitem[Zinnecker \& Yorke(2007)]{zin07} Zinnecker, H., \& Yorke, H.~W.\ 2007, \araa, 45, 481

\bibitem[Zhang et al.(2013)]{zhang13} Zhang, F., Lu, Y., \& Yu, Q.\ 2013, \apj, 768, 153 


\end{thebibliography}
\end{document}